\pdfoutput=1
\documentclass[a4paper, 11pt]{article}

\usepackage[utf8]{inputenc}
\usepackage[T1]{fontenc}
\usepackage[american]{babel}

\usepackage[lining, scale=0.88]{FiraMono}
\usepackage[book, semibold, lining, tabular, scale=0.88]{FiraSans}[2018/05/23]
\usepackage[charter, sfscaled=false, ttscaled=false, euro=false]{mathdesign}
\usepackage[scaled=0.96, lining]{XCharter}
\usepackage{mweights}

\usepackage[sf, bf, raggedright, explicit]{titlesec}
\usepackage[style]{abstract}

\usepackage[max6,affil-it]{authblk}

\usepackage{amsmath}
\usepackage{url,nicefrac,siunitx,xspace,textcomp,color,xcolor,setspace,comment,cite}
\definecolor{darkgreen}{cmyk}{0.9, 0, 0.9, 0}
\definecolor{darkred}{cmyk}{0, 0.6, 0.6, 0}
\usepackage[labelformat=simple,list=true]{subcaption}

\usepackage[
  a4paper,
  width=0.8\paperwidth,
  height=0.76\paperheight,
]{geometry}

\usepackage{etoolbox}
\apptocmd{\thebibliography}{\setlength{\itemsep}{0pt}}{}{}

\usepackage{hyperref}
\hypersetup{
  colorlinks=true,
  linkcolor=blue,   % internal: \ref, \autoref, table of contents
  citecolor=darkgreen,   % \cite
  urlcolor=blue,     % \href and \url
  citebordercolor=0 1 0,
  filebordercolor=0 1 1,
  linkbordercolor=1 0 0,
  menubordercolor=1 0 0,
  urlbordercolor=1 0 0,
  runbordercolor=0 .7 .7,
}
\usepackage{orcidlink}
\usepackage{graphicx}%
\usepackage{multirow}%
\usepackage[title]{appendix}%
\usepackage{textcomp}%
\usepackage{manyfoot}%
\usepackage{booktabs}%
\usepackage{algorithm}%
\usepackage{algorithmicx}%
\usepackage{algpseudocode}%
\usepackage{listings}%

\DeclareGraphicsExtensions{.pdf,.png,.jpg,.eps}
\graphicspath{{./}{./figs/}}

\newcommand{\be}{\begin{equation}}
\newcommand{\ee}{\end{equation}}
\newcommand{\tev}{\mbox{TeV}\xspace}

\newcommand{\papertitle}{MoEDAL-MAPP: the first dedicated search experiment at the LHC}
\title{\papertitle}

\author[1,2]{John R.\ Ellis\,\orcidlink{0000-0002-7399-0813}\thanks{john.ellis@cern.ch} }

\author[1,3]{Nick E.\ Mavromatos\,\orcidlink{0000-0001-7683-8625}\thanks{nikolaos.mavromatos@cern.ch} }

\author[4]{Vasiliki A.\ Mitsou\,\orcidlink{0000-0002-1533-8886}\thanks{Corresponding author. vasiliki.mitsou@ific.uv.es} }

\author[5]{James L.\ Pinfold \orcidlink{0000-0002-2397-4196} \thanks{jpinfold@ualberta.ca}}

\affil[1]{Theoretical Particle Physics and Cosmology Group, Department of Physics, King’s College London, Strand, London WC2R~2LS, United Kingdom}

\affil[2]{Theoretical Physics Department, CERN, CH-1211 Geneva 23, Switzerland}

\affil[3]{Physics Division, School of Applied Mathematical and Physical Sciences, National Technical University of Athens, Zografou Campus, 15780 Athens, Greece}

\affil[4]{Instituto de F\'isica Corpuscular (IFIC), CSIC -- Universitat de Val\`encia, C/ Catedr\'atico Jos\'e Beltr\'an 2, 46980 Paterna (Valencia), Spain}

\affil[5]{Department of Physics, University of Alberta, 4-181 Centre for Interdisciplinary Science, Edmonton, AB~T6G~2E1, Alberta, Canada}

\date{}

\begin{document}

\maketitle

\begin{abstract}
\noindent This special issue brings together a collection of original research articles and review papers addressing topics directly related to the MoEDAL-MAPP experiment at the Large Hadron Collider (LHC). The experiment is designed to search for (meta-)stable, anomalously ionising particles predicted by theoretical extensions of the Standard Model.
Several contributions explore the theoretical and phenomenological aspects of models of particular relevance to the experiment, including scenarios involving magnetic monopoles, highly electrically charged particles, and millicharged particles. Other contributions focus on experimental developments, highlighting novel detection techniques and technological advances aimed at identifying these unconventional signatures and probing new phenomena beyond the Standard Model.
\end{abstract}

\subsubsection*{Editorial on the EPJ Special Topic \newline
\href{https://link.springer.com/collections/ehehgcddfa}{The MoEDAL-MAPP Experiment -- The LHC’s First Dedicated Search Experiment for BSM Physics}}

\hspace*{0.1cm}

%%%%%%%%%%%%%%%%%%%%%%%%%%%%%%%%%%%%%%%%%%%%%%%%%%%%%%%%%%%
%%%%%%%%%%%%%%%%%%%%%%%%%%%%%%%%%%%%%%%%%%%%%%%%%%%%%%%%%%%
\section{Introduction}\label{sec:intro}

MoEDAL (Monopole and Exotics Detector at the LHC), the first dedicated search experiment at the CERN Large Hadron Collider (LHC), is designed to search for exotic, highly ionising particles (HIPs) produced in proton and heavy-ion collisions~\cite{MoEDAL:2014ttp}. Its primary motivation is to pursue the long-standing quest for magnetic monopoles~\cite{Dirac:1931kp,tHooft:1974kcl,Polyakov:1974ek,Mavromatos:2020gwk} and dyons~\cite{Schwinger:1976fr}. Magnetic monopoles provide an explanation for the quantisation of electric charge and restore the symmetry between electric and magnetic fields in Maxwell's equations.
Any massive, (meta-)stable, slow-moving particle carrying a single or multiple units of electric charge may also be detectable by MoEDAL. Such states arise in various scenarios of physics beyond the Standard Model (SM) and may shed light on fundamental questions such as the nature of dark matter and the origin of neutrino masses. Examples include supersymmetric particles, Q-balls, strangelets, and other exotic states~\cite{Fairbairn:2006gg,Alimena:2019zri}.

The detector is deployed around interaction point~8 in the cavern housing the LHCb vertex locator. It is a unique and largely passive detector based on three complementary techniques that do not require electronic readout or trigger. It combines nuclear track detectors (NTDs) with magnetic monopole trapping (MMT) volumes, while the radiation environment is monitored using Timepix detectors.

\begin{figure}[htb]
\centering
  \includegraphics[width=0.75\linewidth]{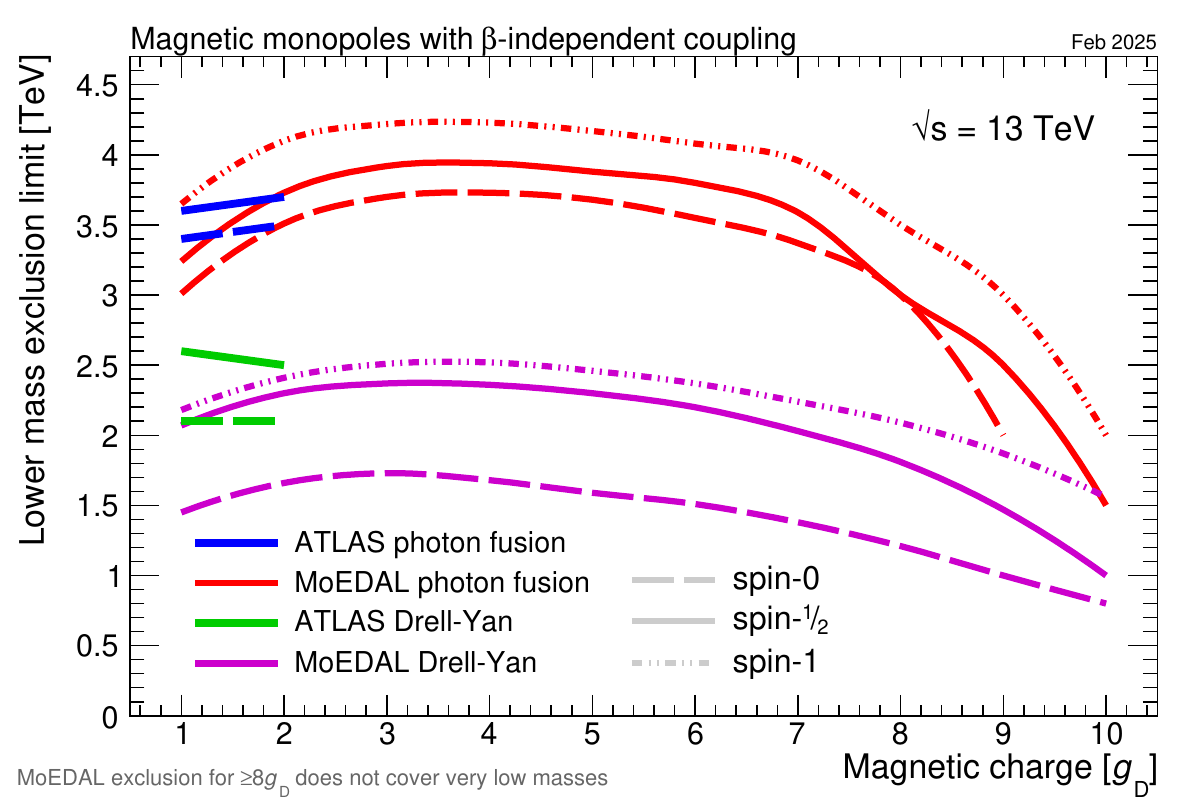}
  \caption{Magnetic-monopole mass limits obtained by MoEDAL~\cite{MoEDAL:2023ost} and ATLAS~\cite{ATLAS:2023esy} using $pp$ collisions at 13~\tev, shown as a function of the magnetic charge. The limits assume Drell--Yan and photon-fusion production with a $\beta$-independent coupling for monopoles of spin~0,\nicefrac{1}{2}, and~1. From\cite{Mitsou:2026zcf}.} \label{fig:atlas-moedal-dy-pf}  
\end{figure}

MoEDAL has pioneered several aspects of magnetic-monopole phenomenology and experimental searches, including studies of vector monopoles~\cite{MoEDAL:2017vhz}, the $\beta$-dependent coupling~\cite{Milton:2006cp,Epele:2012jn}, and photon-fusion production~\cite{Baines:2018ltl,MoEDAL:2019ort,Song:2021vpo}. It performed the first dedicated collider search for dyons~\cite{Schwinger:1976fr,MoEDAL:2020pyb} and remains uniquely positioned to search for particles carrying large magnetic and electric charges~\cite{MoEDAL:2016jlb,MoEDAL:2016lxh,MoEDAL:2017vhz,MoEDAL:2019ort,MoEDAL:2021mpi,MoEDAL:2023ost}, as illustrated in Fig.~\ref{fig:atlas-moedal-dy-pf}.

By considering Schwinger monopole production~\cite{Schwinger:1969ib,Gould:2017zwi,Gould:2019myj,Gould:2021bre,MoEDAL:2021vix,MoEDAL:2024wbc} and employing resummation schemes~\cite{Alexandre:2019iub,Alexandre:2023qjo,Alexandre:2024pbs,Alexandre:2026auj}, MoEDAL researchers have addressed the challenges associated with the large couplings of highly charged particles and derived constraints on composite monopoles.

In addition, the MoEDAL Apparatus for Penetrating Particles (MAPP)\cite{MoEDAL-MAPP:2022kyr} extends the physics reach to feebly interacting, long-lived messengers of physics beyond the Standard Model (BSM), such as millicharged particles\cite{Holdom:1985ag,Holdom:1986eq}. This opens up the exploration of a broad range of new-physics scenarios involving feebly interacting particles, including models involving dark sectors.

This special volume combines original research articles and reviews addressing the physics questions investigated by the MoEDAL-MAPP programme~\cite{MoEDAL-MAPP:2022kyr,Pinfold:2022xnz}, as well as recent advances in the detection techniques employed by the experiment. The contributions are organised into four sections, covering magnetic-monopole theory~\cite{Farakos:2025byy,Mavromatos:2026nlp}, HIP phenomenology~\cite{Maselek:2025cfm,Mitsou:2026ney,Mitsou:2026gvp}, detection techniques~\cite{Mitsou:2026zcf,Kalliokoski:2026yjm,Constantinescu:2026gtq,Bergmann:2026uxe}, and sensitivity to portal models~\cite{deMontigny:2023qft,Kalliokoski:2025oix,Arifeen:2026,Deppisch:2023sga}.

%%%%%%%%%%%%%%%%%%%%%%%%%%%%%%%%%%%%%%%%%%%%%%%%%%%%%%%%%%%%
\section{Magnetic monopole theory}\label{sec:theo}

The theoretical part of the collection of articles in this volume, consisting of two articles~\cite{Farakos:2025byy,Mavromatos:2026nlp}, focuses on whether light electroweak magnetic monopoles, predicted as solutions in theoretical models with masses potentially accessible at colliders, are stable. This is an important question because the stability of such configurations directly affects whether they could persist long enough to be detected experimentally.

Indeed, if a magnetic monopole is unstable, what happens after it is produced at a collider depends on its lifetime and decay channels. If the monopole is very short-lived, it would decay almost immediately into lighter particles and would not reach the detector. In this case, experimental searches would need to focus on its decay products rather than on direct monopole detection. If the monopole is meta-stable, it could travel a measurable distance before decaying, potentially producing a displaced decay vertex or other unusual signatures. If, on the other hand, it is sufficiently long-lived, it could traverse part or all of the detector before decaying, leaving characteristic ionisation or electromagnetic signatures associated with a highly magnetically charged particle. A stable monopole would escape the detector and could potentially produce a particularly distinctive trajectory and energy-loss signature.

The two articles~\cite{Farakos:2025byy,Mavromatos:2026nlp} address different aspects of the theoretical stability of light magnetic monopoles relevant to collider searches, but do not determine their lifetimes or decay rates. Thus, both articles are in the spirit of this volume, which aims to motivate further research into the fascinating field of magnetic monopoles.

Specifically, Farakos \emph{et al.}~\cite{Farakos:2025byy} investigate whether electroweak magnetic monopoles are mechanically stable by examining their energy-momentum tensor, in particular the internal forces and pressure. They apply criteria for mechanical stability inspired by studies in nuclear physics. The analysis emphasises that topological stability and mechanical stability are distinct concepts: a monopole may be topologically protected while still exhibiting mechanical stresses or rotational tendencies. This is particularly relevant for the electroweak Cho--Maison (CM) monopole and its extensions. In the standard CM solution, divergences in the radial and angular components of the internal force signal severe problems with mechanical stability. For the finite-energy Born--Infeld extension of the hypercharge sector, the internal force becomes finite but retains an angular component. Farakos \emph{et al.}  interpret this as indicating that perturbations may induce rotational motion, although this observation alone does not establish an instability.

Mavromatos and Sarkar~\cite{Mavromatos:2026nlp} investigate the stability of the Born--Infeld extension of the CM monopole using a linear perturbation analysis. Exploiting the spherical symmetry of the unperturbed monopole configuration, understood in the sense of invariance under spatial rotations up to gauge transformations, they decompose the perturbations into angular channels and reduce the problem to radial Schr\"odinger/Sturm--Liouville-type eigenvalue equations. The resulting analysis provides evidence in favour of perturbative stability of the Born--Infeld-regularised monopole. However, a complete spectral analysis, which would establish the absence or presence of unstable modes for all channels, is left for future work. Thus, the second article strengthens the case that the Born--Infeld modification can simultaneously render the CM monopole finite-energy and stable against small perturbations, but does not yet provide a rigorous proof of stability for all modes.

%%%%%%%%%%%%%%%%%%%%%%%%%%%%%%%%%%%%%%%%%%%%%%%%%%%%%%%%%%%%
\section{Phenomenology of highly ionsing particles }\label{sec:pheno}

Mase{\l}ek and Sakurai~\cite{Maselek:2025cfm} assess the prospects for detecting electrically charged long-lived particles with the MoEDAL-MAPP experiment. They synthesise findings from recent studies evaluating the sensitivity to supersymmetric models~\cite{Sakurai:2019bac,Felea:2020cvf,Acharya:2020uwc}, radiative neutrino-mass scenarios~\cite{R:2020odv,Hirsch:2021wge}, and generic multiply charged objects~\cite{Altakach:2022hgn}. A key component of this mini-review is a comparative analysis of the reach of MoEDAL relative to the general-purpose ATLAS and CMS experiments. They conclude that, although MoEDAL is limited by its lower integrated luminosity, its passive and effectively background-free detection methodology provides a unique advantage. In particular, the experiment offers important complementarity to the major LHC detectors, especially for signatures involving slow-moving particles and stable states with intermediate electric charges.

The experimental confirmation of non-zero neutrino masses at the sub-eV scale poses a challenge to the otherwise highly successful SM. The EW-$\nu_R$ model, developed by P.~Q.~Hung, addresses this issue through a seesaw mechanism involving non-sterile right-handed neutrinos at the electroweak scale. The model contains mirror fermion doublets and singlets with opposite chirality assignments under the same $SU(2)_W$ gauge symmetry. It also features a viable dark-matter candidate and provides a solution to the strong CP problem. The electroweak-scale mirror quarks and leptons decay into SM quarks and leptons accompanied by very light neutral scalars, resulting in final states containing jets or leptons, scalar mesons, and missing momentum. Mitsou, Staelens, and Hung\cite{Mitsou:2026ney} review the signatures through which these mirror particles can be probed at colliders such as the LHC. Particular emphasis is placed on long-lived mirror particles, whose lifetimes depend on the associated Yukawa couplings and which can give rise to observable displaced leptons and jets, as well as heavy hadrons. The rich spectrum of such states could be accessible to the MoEDAL and MAPP detectors.

In addition to the direct searches for monopoles and highly electrically charged objects (HECOs) carried out by the MoEDAL experiment, the existence of such particles can be probed indirectly through precision measurements. Mitsou and Musumeci~\cite{Mitsou:2026gvp} investigate the contributions of virtual monopoles and HECOs to light-by-light scattering within the frameworks of effective field theories and a Born--Infeld scenario~\cite{Ellis:2017edi,Ellis:2022uxv}, considering the validity regime of these approaches. In particular, measurements of central exclusive photon-pair production with proton tagging, performed by the CMS-TOTEM Precision Proton Spectrometer using LHC proton--proton collision data, are used to constrain the properties of magnetic monopoles and HECOs. In the case of HECOs, their large coupling with the photon and the $Z$ boson is tackled by applying Dyson--Schwinger resummation schemes~\cite{Alexandre:2023qjo,Alexandre:2024pbs}. The resulting limits exclude masses up to a few tens of TeV for monopoles and HECOs with various spins and magnetic and electric charges, respectively.

%%%%%%%%%%%%%%%%%%%%%%%%%%%%%%%%%%%%%%%%%%%%%%%%%%%%%%%%%%%%
\section{Detection techniques}\label{sec:detectors}

MoEDAL relies on passive detectors, such as NTDs~\cite{Kalliokoski:2025sog} and aluminium MMTs~\cite{DeRoeck:2012wua}, to register the passage of highly ionising particles and capture magnetic monopoles, respectively. In addition, Timepix detectors are used to monitor the radiation environment in the cavern~\cite{MOEDAL:2021mrd}.

Mitsou~\cite{Mitsou:2026zcf} reviews the experimental methodologies developed over the years to search for magnetic monopoles and other HIPs. Scintillators, gaseous detectors, and NTDs exploit ionisation and excitation processes, while other approaches rely on induction, Cherenkov radiation, or the catalysis of nucleon decay. The review provides an overview of the relevant experiments and their results, with particular emphasis on the state of the art in searches using cosmic rays and collider experiments, and especially on MoEDAL and ATLAS at the LHC at CERN. The relevance of collider searches for monopoles originating from cosmic sources, together with the resulting experimental constraints, is also discussed. The review concludes with an outlook on future directions in the field.

MoEDAL employs several NTD materials with different detection thresholds in its search for HIPs. The most sensitive material, CR-39, has been used in numerous experiments and is capable of detecting ionisation tracks down to those produced by oxygen ions. However, owing to the high-radiation environment at the LHC, the foils become opaque following chemical etching and exhibit strong scattering, complicating the identification of low-ionisation tracks. Kalliokoski~\cite{Kalliokoski:2026yjm} proposes deploying liquid thin films on both sides of the irradiated foils and applying machine-vision techniques to detect and classify etch pits. The reduction in surface scattering, combined with telecentric imaging, suppresses the LHC-induced haze in the images while simultaneously enhancing the contrast at the edges of the etch pits. This improvement enables more efficient pit detection using machine-vision methods. The transmittance properties of the foils can also be exploited to estimate the overall radiation dose to which they have been exposed.
The improved image contrast further enables the generation of datasets containing relevant features for training neural networks for more detailed foil analysis. However, further studies are required, including investigations of the effects of foil thickness, comparisons between manufacturers, and analyses of the impact of particle incidence angle on etch-pit formation.

Constantinescu \emph{et al.}~\cite{Constantinescu:2026gtq} explore the use of an automated thermal-scanning facility designed to search for molecular defects in NTDs. The system employs a thermal camera mounted on a computer-numerical-control platform to scan the entire surface of the plates under different thermal conditions. The setup provides control over both the plate temperature and the position of the camera along three axes. Custom software has been developed to control the camera movement, heat the pad supporting the NTD plate, and acquire images from the thermal camera. Images acquired using the custom software are compared with those obtained using the camera-specific software, with one sequence covering the entire surface of the plate and another acquired repeatedly over time at the same location to enhance image quality. Preliminary processing of the images obtained with this system shows promising results.

Bergmann \emph{et al.}~\cite{Bergmann:2026uxe} determine the fluences and characteristics of fast neutrons, other hadrons, and HIPs in the MoEDAL environment. These particles may constitute an experimental background for the NTDs used to search for tracks potentially produced by magnetic monopoles or other HIPs. The study is based on data acquired with the Timepix hybrid silicon pixel detector network, the only active detector system installed and operated as part of the MoEDAL experiment. The Timepix detector network enables real-time measurements of mixed radiation fields, including the composition, spectral properties, and directional characteristics of individual radiation components across different regions of the MoEDAL experimental area. Detailed results of the radiation-field analysis are presented, with particular emphasis on neutrons and HIPs, including their directional distributions. The first results demonstrating the spatial tracking capabilities of the Timepix detectors are also reported, illustrating the reconstruction of particle directions and energy-loss profiles from individual detector frames.

%%%%%%%%%%%%%%%%%%%%%%%%%%%%%%%%%%%%%%%%%%%%%%%%%%%%%%%%%%%%
\section{Sensitivity to portal models}\label{sec:portal}

Millicharged particles (mCPs), hypothetical free particles with effective electric charges much smaller than the elementary charge, $e$, offer a valuable probe of dark sectors and fundamental physics through several clear experimental signatures. Various models of BSM  predict such particles, the existence of which could help elucidate the ongoing mysteries regarding electric charge quantisation and the nature of dark matter. Moreover, a hypothetical scenario involving a small millicharged subcomponent of dark matter has recently been demonstrated as a viable explanation of the anomaly in the 21~cm hydrogen absorption signal reported by the EDGES Collaboration~\cite{Munoz:2018pzp,Berlin:2018sjs,Kovetz:2018zan}. Although several decades of indirect observations and direct experimental searches for mCPs at particle accelerators have led to severe constraints, a substantial window of the mCP mass-mixing parameter space remains unexplored at the energy frontier accessible to current state-of-the-art accelerators, such as the LHC. Consequently, mCPs have remained topical over the years, and new experimental searches at accelerators have been gaining interest. De Montigny \emph{et al.}~\cite{deMontigny:2023qft} review the theoretical frameworks in which mCPs emerge and their phenomenological implications, the current direct and indirect constraints on mCPs, and the present state of the ongoing and upcoming searches for mCPs at particle accelerators.

Kalliokoski \emph{et al.}~\cite{Kalliokoski:2025oix} present a detailed study of the projected background-free sensitivity of the MAPP Outrigger Detector (OD) to mCPs at the High-Luminosity LHC (HL-LHC). As the first upgrade to the MAPP experiment~\cite{Kalliokoski:2023cgw}, the MAPP OD is a standalone detector designed to offer enhanced sensitivity to high-mass mCPs with intermediate effective charges. The MAPP OD is planned for installation in a duct adjacent to the MAPP-1 detector, located between the LHC’s UA83 gallery and the beamline. Considering mCP production via the Drell--Yan mechanism and various meson decays, the results show that, at the 95\% confidence level, the MAPP OD can extend the experiment’s upper mass reach to mCP masses of approximately 200~GeV at the HL-LHC.

Arifeen \emph{et al.}~\cite{Arifeen:2026} present a comprehensive calculation of the ionisation energy loss for relativistic neutral particles carrying electric dipole moments (EDMs). The stopping power formula is derived explicitly for the aligned configuration and it is discussed how general dipole orientation modifies the result using known differential cross-section structure. The results show that the energy loss scales differently from charged particles due to the $r^{-3}$ dipole field behaviour, yielding no logarithmic enhancement. For a 100-GeV heavy neutrino with an EDM of $10^{-16}~e\cdot\mathrm{cm}$, ionisation signals of order $10~\mathrm{keV/m}$ are predicted for $\gamma\sim2.3$, approximately $10^{-3}$ that of a minimally ionising charged particle. Currently, such particles could, in principle, be detected by the MAPP-1 detector ~\cite{Kalliokoski:2023cgw,Kalliokoski:2025oix}.

Deppisch, Kulkarni and Liu~\cite{Deppisch:2023sga} investigate the possibility of searching for right-handed neutrinos at the MAPP detector. In particular, pair-production of right-handed neutrinos $N$ from either a $B-L$ gauge boson $Z'$~\cite{Deppisch:2019kvs}, as well as a SM $Z$ boson are considered. Under a no-background assumption, it is  shown that the MAPP detector can be sensitive to active-sterile neutrino mixing strengths as low as $V_{\mu N}^2 \approx 10^{-12}$ for multiple choices of $m_N/m_{Z'}$ values, when taking the $B-L$ gauge coupling $g_{B-L}=10^{-3}$ near its current limit. The SM $Z$ boson portal can reach a similar sensitivity, when the effective mixing between the $B-L$ and SM gauge boson is $\alpha \approx 0.002$.

%%%%%

%%%%%%%%%%%%%%%%%%%%%%%%%%%%%%%%%%%%%%%%%%%%%%%%%%%%%%%%%%%%
\section{Final words}\label{sec:summary}

The LHC is reaching collision energies never before achieved at particle colliders, opening new opportunities for the discovery of exotic particles in the TeV mass range.
Within this landscape, MoEDAL is currently the only experiment specifically designed to detect signatures of (meta-)stable, highly ionising particles. The MAPP extension aims to broaden this discovery potential by extending the search to feebly interacting particles.

This special issue comprises 13 contributions highlighting recent theoretical, phenomenological, and experimental advances in areas closely related to the MoEDAL-MAPP programme. Beyond the exploration of largely uncharted territory it has achieved, the MoEDAL-MAPP experiment has also helped to revitalise specific areas of new physics beyond the Standard Model, underscoring its broader impact on the field.

%%%%%%%%%%%%%%%%%%%%%%%%%%%%%%%%%%%%%%%%%%%%%%%%
\subsection*{Acknowledgements}
The editors of this special issue wish to thank the authors for their high-quality contributions and the article reviewers for their insightful and critical comments.
We are grateful to the editorial team of the European
Physical Journal Special Topics for their patience, support and guidance throughout the publication process, in particular to Sandrine Karpe and Lathapriya Balamurugan.
The work of NEM is supported in part by the Science and Technology Facilities Council (UK) under
grant ST/X000753/1.
VAM acknowledges support by MICIN/AEI/10.13039/501100011033/ FEDER, EU via the grant  PID2024-158190NB-C21 and by MICIU/AEI grant Severo Ochoa CEX2023-001292-S. 

%%%%%%%%%%%%%%%%%%%%%%%%%%%%%%%%%%%%%%%%%%%%%%%%
\subsection*{Data Availability}
No new or original datasets were created in this study.

%%%%%%%%%%%%%%%%%%%%%%%%%%%%%%%%%%%%%%%%%%%%%%%%

%\bibliographystyle{jhep.bst}
\bibliographystyle{JHEP-vaso}
\bibliography{biblio}

@article{Farakos:2025byy,
    author = "Farakos, Konstantinos and Koutsoumbas, George and Mavromatos, Nick E. and Zarafonitis, Alexandros",
    title = "{On internal mechanical properties of electroweak magnetic monopoles and their effects on stability}",
    eprint = "2506.04872",
    archivePrefix = "arXiv",
    primaryClass = "hep-th",
    reportNumber = "KCL-PH-TH/2025-20",
    doi = "10.1140/epjs/s11734-025-02083-z",
    journal = "Eur. Phys. J. ST",
    volume = "235",
    number = "25",
    pages = "5361--5413",
    year = "2026",
    note = "[Erratum: Eur. Phys. J. ST 235, 5415 (2026)]"
}

@article{Mavromatos:2026nlp,
    author = "Mavromatos, N. E. and Sarkar, Sarben",
    title = "{On the stability of Born--Infeld-regularised electroweak monopoles}",
    eprint = "2602.01921",
    archivePrefix = "arXiv",
    primaryClass = "hep-th",
    reportNumber = "KCL-PH-TH/2026- 2",
    doi = "10.1140/epjs/s11734-026-02268-0",
    journal = "Eur. Phys. J. ST",
    volume = "235",
    number = "25",
    pages = "5417--5442",
    year = "2026"
}

@article{Maselek:2025cfm,
    author = "Mase{\l}ek, Rafa{\l} and Sakurai, Kazuki",
    title = "{Prospects for detecting charged long-lived BSM particles at MoEDAL-MAPP experiment: a mini-review}",
    eprint = "2512.23387",
    archivePrefix = "arXiv",
    primaryClass = "hep-ph",
    doi = "10.1140/epjs/s11734-026-02182-5",
    journal = "Eur. Phys. J. ST",
    volume = "235",
    number = "25",
    pages = "5443--5452",
    year = "2026"
}

@article{Mitsou:2026ney,
    author = "Mitsou, Vasiliki A. and Staelens, Michael and Hung, Pham Quang",
    title = "{Signatures and probes of mirror mesons at colliders}",
    doi = "10.1140/epjs/s11734-026-02498-2",
    journal = "Eur. Phys. J. ST",
    volume = "235",
    number = "25",
    pages = "5453--5467",
    year = "2026"
}

@article{Mitsou:2026gvp,
    author = "Mitsou, Vasiliki A. and Musumeci, Emanuela",
    title = "{Constraining magnetic monopoles and multiply charged particles with diphoton events at the LHC}",
    eprint = "2604.07300",
    archivePrefix = "arXiv",
    primaryClass = "hep-ph",
    doi = "10.1140/epjs/s11734-026-02550-1",
    journal = "Eur. Phys. J. ST",
    volume = "235",
    number = "25",
    pages = "5469--5480",
    year = "2026"
}

@article{Mitsou:2026zcf,
    author = "Mitsou, Vasiliki A.",
    title = "{Magnetic monopoles: from Dirac to the Large Hadron Collider}",
    eprint = "2605.02869",
    archivePrefix = "arXiv",
    primaryClass = "hep-ex",
    doi = "10.1140/epjs/s11734-026-02463-z",
    journal = "Eur. Phys. J. ST",
    volume = "235",
    number = "25",
    pages = "5481--5521",
    year = "2026"
}

@article{Kalliokoski:2026yjm,
    author = "Kalliokoski, Matti",
    title = "{Increased sensitivity in optical scanning of irradiated CR39 foils with liquid thin films}",
    doi = "10.1140/epjs/s11734-026-02266-2",
    journal = "Eur. Phys. J. ST",
    volume = "235",
    number = "25",
    pages = "5523--5530",
    year = "2026"
}

@article{Constantinescu:2026gtq,
    author = "Constantinescu, Vlad and Br{\^a}nzaș, Horea and Daniel, Felea and Tordai, Gavril{\u{a}}",
    title = "{Automated thermal scanning installation of nuclear track detector plates: design and preliminary results}",
    doi = "10.1140/epjs/s11734-025-02120-x",
    journal = "Eur. Phys. J. ST",
    volume = "235",
    number = "25",
    pages = "5531--5540",
    year = "2026"
}

@article{Bergmann:2026uxe,
    author = "Bergmann, Benedikt and Burian, Petr and Jane{\v{c}}ek, Josef and Leroy, Claude and M{\'a}nek, Petr and Pinfold, James and Posp{\'\i}{\v{s}}il, Stanislav and Soluk, Richard and Suk, Michal",
    title = "{Analysis of mixed radiation fields at the MoEDAL experiment based on real-time data from a Timepix detector network}",
    eprint = "2605.05936",
    archivePrefix = "arXiv",
    primaryClass = "physics.ins-det",
    doi = "10.1140/epjs/s11734-026-02586-3",
    journal = "Eur. Phys. J. ST",
    volume = "235",
    number = "25",
    pages = "5541--5553",
    year = "2026"
}

@article{deMontigny:2023qft,
    author = "de Montigny, Marc and Ouimet, Pierre-Philippe A. and Pinfold, James and Shaa, Ameir and Staelens, Michael",
    title = "{Minicharged particles at accelerators: progress and prospects}",
    eprint = "2307.07855",
    archivePrefix = "arXiv",
    primaryClass = "hep-ph",
    doi = "10.1140/epjs/s11734-026-02362-3",
    journal = "Eur. Phys. J. ST",
    volume = "235",
    number = "25",
    pages = "5555--5571",
    year = "2026"
}

@article{Kalliokoski:2025oix,
    author = "Kalliokoski, Matti and Mitsou, Vasiliki A. and de Montigny, Marc and Mukhopadhyay, Abhinab and Ouimet, Pierre-Philippe A. and Pinfold, James and Shaa, Ameir and Staelens, Michael",
    title = "{Minicharged particle sensitivity of the MAPP Outrigger Detector}",
    eprint = "2508.04896",
    archivePrefix = "arXiv",
    primaryClass = "hep-ph",
    doi = "10.1140/epjs/s11734-025-01888-2",
    journal = "Eur. Phys. J. ST",
    volume = "235",
    number = "25",
    pages = "5573--5582",
    year = "2026"
}

@article{Arifeen:2026,
    author = "Arifeen, Shafakat and Mukhopadhyay, Abhinab and Ouimet, Pierre-Philippe and Pinfold, James Lewis and Shaa, Ameir and Staelens, Michael",
    title = "{Ionization energy loss of relativistic particles with electric dipole moments}",
    doi = "10.1140/epjs/s11734-026-02374-z",
    journal = "Eur. Phys. J. ST",
    volume = "235",
    number = "25",
    pages = "5583--5594",
    year = "2026"
}

@article{Deppisch:2023sga,
    author = "Deppisch, Frank F. and Kulkarni, Suchita and Liu, Wei",
    title = "{Sterile neutrinos at MAPP in the $B-L$ model}",
    eprint = "2311.01719",
    archivePrefix = "arXiv",
    primaryClass = "hep-ph",
    doi = "10.1140/epjs/s11734-026-02205-1",
    journal = "Eur. Phys. J. ST",
    volume = "235",
    number = "25",
    pages = "5595--5606",
    year = "2026"
}

@article{MoEDAL:2014ttp,
    author = "Acharya, B. and others",
    collaboration = "MoEDAL",
    title = "{The physics programme of the MoEDAL experiment at the LHC}",
    eprint = "1405.7662",
    archivePrefix = "arXiv",
    primaryClass = "hep-ph",
    reportNumber = "KCL-PH-TH-2014-02, LCTS-2014-02, CERN-PH-TH-2014-021, IFIC-14-16, IMPERIAL-TP-2014-AR-1, KCL-PH-TH/2014-02, LCTS/2014-02, CERN-PH-TH/2014-021, IFIC/14-16,
  Imperial/TP/2014/AR/1",
    doi = "10.1142/S0217751X14300506",
    journal = "Int. J. Mod. Phys. A",
    volume = "29",
    pages = "1430050",
    year = "2014"
}

@article{MoEDAL:2016jlb,
    author = "Acharya, B. and others",
    collaboration = "MoEDAL",
    title = "{Search for magnetic monopoles with the MoEDAL prototype trapping detector in 8 TeV proton--proton collisions at the LHC}",
    eprint = "1604.06645",
    archivePrefix = "arXiv",
    primaryClass = "hep-ex",
    reportNumber = "CERN-EP-2016-101",
    doi = "10.1007/JHEP08(2016)067",
    journal = "JHEP",
    volume = "08",
    pages = "067",
    year = "2016"
}

@article{MoEDAL:2016lxh,
    author = "Acharya, B. and others",
    collaboration = "MoEDAL",
    title = "{Search for magnetic monopoles with the MoEDAL forward trapping detector in 13 TeV proton--proton collisions at the LHC}",
    eprint = "1611.06817",
    archivePrefix = "arXiv",
    primaryClass = "hep-ex",
    reportNumber = "CERN-EP-2016-282",
    doi = "10.1103/PhysRevLett.118.061801",
    journal = "Phys. Rev. Lett.",
    volume = "118",
    number = "6",
    pages = "061801",
    year = "2017"
}

@article{MoEDAL:2017vhz,
    author = "Acharya, B. and others",
    collaboration = "MoEDAL",
    title = "{Search for magnetic monopoles with the MoEDAL forward trapping detector in 2.11 fb$^{-1}$ of 13 TeV proton--proton collisions at the LHC}",
    eprint = "1712.09849",
    archivePrefix = "arXiv",
    primaryClass = "hep-ex",
    reportNumber = "CERN-EP-2017-342",
    doi = "10.1016/j.physletb.2018.05.069",
    journal = "Phys. Lett. B",
    volume = "782",
    pages = "510--516",
    year = "2018"
}

@article{MoEDAL:2019ort,
    author = "Acharya, B. and others",
    collaboration = "MoEDAL",
    title = "{Magnetic monopole search with the full MoEDAL trapping detector in 13~TeV $pp$ collisions interpreted in photon-fusion and Drell--Yan production}",
    eprint = "1903.08491",
    archivePrefix = "arXiv",
    primaryClass = "hep-ex",
    reportNumber = "CERN-EP-2019-039, IFIC/19-09, KCL-PH-TH/2019-17",
    doi = "10.1103/PhysRevLett.123.021802",
    journal = "Phys. Rev. Lett.",
    volume = "123",
    number = "2",
    pages = "021802",
    year = "2019"
}

@article{MoEDAL:2020pyb,
    author = "Acharya, B. and others",
    collaboration = "MoEDAL",
    title = "{First search for dyons with the full MoEDAL trapping detector in 13~TeV $pp$ collisions}",
    eprint = "2002.00861",
    archivePrefix = "arXiv",
    primaryClass = "hep-ex",
    doi = "10.1103/PhysRevLett.126.071801",
    journal = "Phys. Rev. Lett.",
    volume = "126",
    number = "7",
    pages = "071801",
    year = "2021"
}

@article{MoEDAL:2021vix,
    author = "Acharya, B. and others",
    collaboration = "MoEDAL",
    title = "{Search for magnetic monopoles produced via the Schwinger mechanism}",
    eprint = "2106.11933",
    archivePrefix = "arXiv",
    primaryClass = "hep-ex",
    doi = "10.1038/s41586-021-04298-1",
    journal = "Nature",
    volume = "602",
    number = "7895",
    pages = "63--67",
    year = "2022"
}

@article{MoEDAL:2021mpi,
    author = "Acharya, B. and others",
    collaboration = "MoEDAL",
    title = "{Search for highly-ionizing particles in $pp$ collisions at the LHC\textquoteright{}s Run-1 using the prototype MoEDAL detector}",
    eprint = "2112.05806",
    archivePrefix = "arXiv",
    primaryClass = "hep-ex",
    doi = "10.1140/epjc/s10052-022-10608-2",
    journal = "Eur. Phys. J. C",
    volume = "82",
    number = "8",
    pages = "694",
    year = "2022"
}

@article{MOEDAL:2021mrd,
    author = "Bergmann, Benedikt and others",
    collaboration = "MoEDAL",
    title = "{Timepix3 as solid-state time-projection chamber in particle and nuclear physics}",
    doi = "10.22323/1.390.0720",
    journal = "PoS",
    volume = "ICHEP2020",
    pages = "720",
    year = "2021"
}

@inproceedings{MoEDAL-MAPP:2022kyr,
    author = "Acharya, B. and others",
    collaboration = "MoEDAL-MAPP",
    title = "{MoEDAL-MAPP, an LHC dedicated detector search facility}",
    booktitle = "{Snowmass 2021}",
    eprint = "2209.03988",
    archivePrefix = "arXiv",
    primaryClass = "hep-ph",
    month = "9",
    year = "2022"
}

@article{MoEDAL:2024wbc,
    author = "Acharya, B. and others",
    collaboration = "MoEDAL",
    title = "{MoEDAL search in the CMS beam pipe for magnetic monopoles produced via the Schwinger effect}",
    eprint = "2402.15682",
    archivePrefix = "arXiv",
    primaryClass = "nucl-ex",
    doi = "10.1103/PhysRevLett.133.071803",
    journal = "Phys. Rev. Lett.",
    volume = "133",
    number = "7",
    pages = "071803",
    year = "2024"
}

@article{MoEDAL:2023ost,
    author = "Acharya, B. and others",
    collaboration = "MoEDAL",
    title = "{Search for highly ionizing particles in $pp$ collisions during LHC Run 2 using the full MoEDAL detector}",
    eprint = "2311.06509",
    archivePrefix = "arXiv",
    primaryClass = "hep-ex",
    doi = "10.1103/PhysRevLett.134.071802",
    journal = "Phys. Rev. Lett.",
    volume = "134",
    number = "7",
    pages = "071802",
    year = "2025"
}

@article{Dirac:1931kp,
    author = "Dirac, Paul Adrien Maurice",
    title = "{Quantised singularities in the electromagnetic field,}",
    reportNumber = "RX-722",
    doi = "10.1098/rspa.1931.0130",
    journal = "Proc. Roy. Soc. Lond. A",
    volume = "133",
    number = "821",
    pages = "60--72",
    year = "1931"
}

@article{tHooft:1974kcl,
    author = "'t Hooft, Gerard",
    editor = "Taylor, J. C.",
    title = "{Magnetic monopoles in unified gauge theories}",
    reportNumber = "CERN-TH-1876",
    doi = "10.1016/0550-3213(74)90486-6",
    journal = "Nucl. Phys. B",
    volume = "79",
    pages = "276--284",
    year = "1974"
}

@article{Polyakov:1974ek,
    author = "Polyakov, Alexander M.",
    editor = "Taylor, J. C.",
    title = "{Particle spectrum in quantum field theory}",
    reportNumber = "PRINT-74-1566 (LANDAU-INST)",
    journal = "JETP Lett.",
    volume = "20",
    pages = "194--195",
    year = "1974"
}

@article{Schwinger:1976fr,
    author = "Schwinger, Julian S. and Milton, Kimball A. and Tsai, Wu-yang and DeRaad, Jr., Lester L. and Clark, Donald C.",
    editor = "Milton, K. A.",
    title = "{Nonrelativistic dyon\textendash{}dyon scattering}",
    reportNumber = "UCLA/76/TEP/3",
    doi = "10.1016/0003-4916(76)90020-8",
    journal = "Annals Phys.",
    volume = "101",
    pages = "451",
    year = "1976"
}

@article{Schwinger:1969ib,
    author = "Schwinger, Julian S.",
    title = "{A magnetic model of matter}",
    doi = "10.1126/science.165.3895.757",
    journal = "Science",
    volume = "165",
    pages = "757--761",
    year = "1969"
}

@article{Gould:2021bre,
    author = "Gould, Oliver and Ho, David L. -J. and Rajantie, Arttu",
    title = "{Schwinger pair production of magnetic monopoles: momentum distribution for heavy-ion collisions}",
    eprint = "2103.14454",
    archivePrefix = "arXiv",
    primaryClass = "hep-ph",
    reportNumber = "HIP-2021-12/TH, IMPERIAL-TP-2021-DH-05",
    doi = "10.1103/PhysRevD.104.015033",
    journal = "Phys. Rev. D",
    volume = "104",
    number = "1",
    pages = "015033",
    year = "2021"
}

@article{Gould:2019myj,
    author = "Gould, Oliver and Ho, David L. -J. and Rajantie, Arttu",
    title = "{Towards Schwinger production of magnetic monopoles in heavy-ion collisions}",
    eprint = "1902.04388",
    archivePrefix = "arXiv",
    primaryClass = "hep-th",
    reportNumber = "IMPERIAL-TP-2019-DH-01, HIP-2019-2/TH, IMPERIAL-TP-2019-DH-01; HIP-2019-2/TH",
    doi = "10.1103/PhysRevD.100.015041",
    journal = "Phys. Rev. D",
    volume = "100",
    number = "1",
    pages = "015041",
    year = "2019"
}

@article{Gould:2017zwi,
    author = "Gould, Oliver and Rajantie, Arttu",
    title = "{Magnetic monopole mass bounds from heavy ion collisions and neutron stars}",
    eprint = "1705.07052",
    archivePrefix = "arXiv",
    primaryClass = "hep-ph",
    reportNumber = "IMPERIAL-TP-2017-OG-2",
    doi = "10.1103/PhysRevLett.119.241601",
    journal = "Phys. Rev. Lett.",
    volume = "119",
    number = "24",
    pages = "241601",
    year = "2017"
}

@article{Baines:2018ltl,
    author = "Baines, S. and Mavromatos, N. E. and Mitsou, V. A. and Pinfold, J. L. and Santra, A.",
    title = "{Monopole production via photon fusion and Drell\textendash{}Yan processes: MadGraph implementation and perturbativity via velocity-dependent coupling and magnetic moment as novel features}",
    eprint = "1808.08942",
    archivePrefix = "arXiv",
    primaryClass = "hep-ph",
    reportNumber = "KCL-PH-TH/2018-43, IFIC/18-32",
    doi = "10.1140/epjc/s10052-018-6440-6",
    journal = "Eur. Phys. J. C",
    volume = "78",
    number = "11",
    pages = "966",
    year = "2018"
}

@article{Song:2021vpo,
    author = "Song, Wen Yi and Taylor, Wendy",
    title = "{Pair production of magnetic monopoles and stable high-electric-charge objects in proton{\textendash}proton and heavy-ion collisions}",
    eprint = "2107.10789",
    archivePrefix = "arXiv",
    primaryClass = "hep-ph",
    doi = "10.1088/1361-6471/ac3dce",
    journal = "J. Phys. G",
    volume = "49",
    number = "4",
    pages = "045002",
    year = "2022"
}

@article{Mavromatos:2020gwk,
    author = "Mavromatos, Nick E. and Mitsou, Vasiliki A.",
    title = "{Magnetic monopoles revisited: models and searches at colliders and in the Cosmos}",
    eprint = "2005.05100",
    archivePrefix = "arXiv",
    primaryClass = "hep-ph",
    reportNumber = "KCL-PH-TH/2019-94, IFIC/20-12",
    doi = "10.1142/S0217751X20300124",
    journal = "Int. J. Mod. Phys. A",
    volume = "35",
    number = "23",
    pages = "2030012",
    year = "2020"
}

@article{Fairbairn:2006gg,
    author = "Fairbairn, M. and Kraan, A. C. and Milstead, D. A. and Sjostrand, T. and Skands, Peter Z. and Sloan, T.",
    title = "{Stable massive particles at colliders}",
    eprint = "hep-ph/0611040",
    archivePrefix = "arXiv",
    reportNumber = "FERMILAB-PUB-07-034-T",
    doi = "10.1016/j.physrep.2006.10.002",
    journal = "Phys. Rept.",
    volume = "438",
    pages = "1--63",
    year = "2007"
}

@article{Alimena:2019zri,
    author = "Alimena, Juliette and others",
    title = "{Searching for long-lived particles beyond the Standard Model at the Large Hadron Collider}",
    eprint = "1903.04497",
    archivePrefix = "arXiv",
    primaryClass = "hep-ex",
    doi = "10.1088/1361-6471/ab4574",
    journal = "J. Phys. G",
    volume = "47",
    number = "9",
    pages = "090501",
    year = "2020"
}

@article{Milton:2006cp,
    author = "Milton, Kimball A.",
    title = "{Theoretical and experimental status of magnetic monopoles}",
    eprint = "hep-ex/0602040",
    archivePrefix = "arXiv",
    doi = "10.1088/0034-4885/69/6/R02",
    journal = "Rept. Prog. Phys.",
    volume = "69",
    pages = "1637--1712",
    year = "2006"
}

@article{Epele:2012jn,
    author = "Epele, Luis N. and Fanchiotti, Huner and Canal, Carlos A. Garcia and Mitsou, Vasiliki A. and Vento, Vicente",
    title = "{Looking for magnetic monopoles at LHC with diphoton events}",
    eprint = "1205.6120",
    archivePrefix = "arXiv",
    primaryClass = "hep-ph",
    doi = "10.1140/epjp/i2012-12060-8",
    journal = "Eur. Phys. J. Plus",
    volume = "127",
    pages = "60",
    year = "2012"
}

@article{Alexandre:2019iub,
    author = "Alexandre, Jean and Mavromatos, Nick E.",
    title = "{Weak-U(1) \texttimes{} strong-U(1) effective gauge field theories and electron--monopole scattering}",
    eprint = "1906.08738",
    archivePrefix = "arXiv",
    primaryClass = "hep-ph",
    reportNumber = "KCL-PH-TH/2019-46",
    doi = "10.1103/PhysRevD.100.096005",
    journal = "Phys. Rev. D",
    volume = "100",
    number = "9",
    pages = "096005",
    year = "2019"
}

@article{Alexandre:2023qjo,
    author = "Alexandre, Jean and Mavromatos, Nick E. and Mitsou, Vasiliki A. and Musumeci, Emanuela",
    title = "{Resummation schemes for high-electric-charge objects leading to improved experimental mass limits}",
    eprint = "2310.17452",
    archivePrefix = "arXiv",
    primaryClass = "hep-ph",
    reportNumber = "KCL-PH-TH/2023-36",
    doi = "10.1103/PhysRevD.109.036026",
    journal = "Phys. Rev. D",
    volume = "109",
    number = "3",
    pages = "036026",
    year = "2024"
}

@article{Alexandre:2024pbs,
    author = "Alexandre, Jean and Mavromatos, Nick E. and Mitsou, Vasiliki A. and Musumeci, Emanuela",
    title = "{Impact of resummation on the production and experimental bounds of scalar high-electric-charge objects}",
    eprint = "2412.19001",
    archivePrefix = "arXiv",
    primaryClass = "hep-ph",
    reportNumber = "KCL-PH-TH/2024-35",
    doi = "10.1103/PhysRevD.111.076010",
    journal = "Phys. Rev. D",
    volume = "111",
    number = "7",
    pages = "076010",
    year = "2025"
}

@article{Alexandre:2026auj,
    author = "Alexandre, Jean and Mavromatos, Nick E. and Mitsou, Vasiliki A. and Musumeci, Emanuela",
    title = "{Magnetic-monopole resummation justifies perturbatively calculated collider production cross sections}",
    eprint = "2603.24200",
    archivePrefix = "arXiv",
    primaryClass = "hep-ph",
    reportNumber = "KCL-PH-TH/2025-45",
    doi = "10.1140/epjc/s10052-026-16085-1",
    journal = "Eur. Phys. J. C",
    volume = "86",
    number = "8",
    pages = "942",
    year = "2026"
}

@article{Holdom:1985ag,
    author = "Holdom, Bob",
    title = "{Two U(1)'s and epsilon charge shifts}",
    reportNumber = "UTPT-85-30",
    doi = "10.1016/0370-2693(86)91377-8",
    journal = "Phys. Lett. B",
    volume = "166",
    pages = "196--198",
    year = "1986"
}

@article{Holdom:1986eq,
    author = "Holdom, Bob",
    title = "{Searching for $\epsilon$ charges and a new U(1)}",
    reportNumber = "UTPT-86-03",
    doi = "10.1016/0370-2693(86)90470-3",
    journal = "Phys. Lett. B",
    volume = "178",
    pages = "65--70",
    year = "1986"
}

@article{Kalliokoski:2023cgw,
    author = "Kalliokoski, Matti and Mitsou, Vasiliki A. and de Montigny, Marc and Mukhopadhyay, Abhinab and Ouimet, Pierre-Philippe A. and Pinfold, James and Shaa, Ameir and Staelens, Michael",
    title = "{Searching for minicharged particles at the energy frontier with the MoEDAL-MAPP experiment at the LHC}",
    eprint = "2311.02185",
    archivePrefix = "arXiv",
    primaryClass = "hep-ph",
    doi = "10.1007/JHEP04(2024)137",
    journal = "JHEP",
    volume = "04",
    pages = "137",
    year = "2024"
}

@article{Sakurai:2019bac,
    author = "Sakurai, K. and Felea, D. and Mamuzic, J. and Mavromatos, N. E. and Mitsou, V. A. and Pinfold, J. L. and Ruiz de Austri, R. and Santra, A. and Vives, O.",
    editor = "Widmann, Eberhard and Marton, Johann and Pichler, Andreas and Simon, Martin and Murtagh, Daniel",
    title = "{SUSY discovery prospects with MoEDAL}",
    eprint = "1903.11022",
    archivePrefix = "arXiv",
    primaryClass = "hep-ph",
    reportNumber = "IFIC/19-16, KCL-PH-TH/2019-34",
    doi = "10.1088/1742-6596/1586/1/012018",
    journal = "J. Phys. Conf. Ser.",
    volume = "1586",
    number = "1",
    pages = "012018",
    year = "2020"
}

@article{Felea:2020cvf,
    author = "Felea, D. and Mamuzic, J. and Mase\l{}ek, R. and Mavromatos, N. E. and Mitsou, V. A. and Pinfold, J. L. and Ruiz de Austri, R. and Sakurai, K. and Santra, A. and Vives, O.",
    title = "{Prospects for discovering supersymmetric long-lived particles with MoEDAL}",
    eprint = "2001.05980",
    archivePrefix = "arXiv",
    primaryClass = "hep-ph",
    reportNumber = "IFIC/19-58, FTUV-19-1216, KCL-PH-TH/2019-93",
    doi = "10.1140/epjc/s10052-020-7994-7",
    journal = "Eur. Phys. J. C",
    volume = "80",
    number = "5",
    pages = "431",
    year = "2020"
}

@article{Acharya:2020uwc,
    author = "Acharya, B. S. and De Roeck, A. and Ellis, J. and Ghosh, D. K. and Mase\l{}ek, R. and Panizzo, G. and Pinfold, J. L. and Sakurai, K. and Shaa, A. and Wall, A.",
    title = "{Prospects of searches for long-lived charged particles with MoEDAL}",
    eprint = "2004.11305",
    archivePrefix = "arXiv",
    primaryClass = "hep-ph",
    reportNumber = "KCL-PH-TH/2020-18, CERN-TH-2020-060",
    doi = "10.1140/epjc/s10052-020-8093-5",
    journal = "Eur. Phys. J. C",
    volume = "80",
    number = "6",
    pages = "572",
    year = "2020"
}

@article{R:2020odv,
    author = "Arbel\'aez, Carolina and Cottin, Giovanna and Helo, Juan Carlos and Hirsch, Martin",
    title = "{Long-lived charged particles and multi-lepton signatures from neutrino mass models}",
    eprint = "2003.11494",
    archivePrefix = "arXiv",
    primaryClass = "hep-ph",
    doi = "10.1103/PhysRevD.101.095033",
    journal = "Phys. Rev. D",
    volume = "101",
    number = "9",
    pages = "095033",
    year = "2020"
}

@article{Hirsch:2021wge,
    author = "Hirsch, Martin and Mase\l{}ek, Rafa\l{} and Sakurai, Kazuki",
    title = "{Detecting long-lived multi-charged particles in neutrino mass models with MoEDAL}",
    eprint = "2103.05644",
    archivePrefix = "arXiv",
    primaryClass = "hep-ph",
    doi = "10.1140/epjc/s10052-021-09507-9",
    journal = "Eur. Phys. J. C",
    volume = "81",
    number = "8",
    pages = "697",
    year = "2021"
}

@article{Altakach:2022hgn,
    author = "Altakach, Mohammad Mahdi and Lamba, Priyanka and Mase\l{}ek, Rafa\l{} and Mitsou, Vasiliki A. and Sakurai, Kazuki",
    title = "{Discovery prospects for long-lived multiply charged particles at the LHC}",
    eprint = "2204.03667",
    archivePrefix = "arXiv",
    primaryClass = "hep-ph",    
    reportNumber = "IFIC/22-13",
    doi = "10.1140/epjc/s10052-022-10805-z",
    journal = "Eur. Phys. J. C",
    volume = "82",
    number = "9",
    pages = "848",
    year = "2022"
}

@article{Pinfold:2022xnz,
    author = "Pinfold, James",
    title = "{The MoEDAL-MAPP experiment at the LHC {\textendash} the continuation of the LHC{\textquoteright}s 1st dedicated search experiment}",
    doi = "10.22323/1.414.0168",
    journal = "PoS",
    volume = "ICHEP2022",
    pages = "168",
    month = "11",
    year = "2022"
}

@article{Kalliokoski:2025sog,
    author = "Kalliokoski, M. and others",
    title = "{Calibration of solid state nuclear track detectors for rare event searches}",
    eprint = "2503.09495",
    archivePrefix = "arXiv",
    primaryClass = "physics.ins-det",
    doi = "10.1088/1748-0221/20/03/P03014",
    journal = "JINST",
    volume = "20",
    number = "03",
    pages = "P03014",
    year = "2025"    
}

@article{DeRoeck:2012wua,
    author = {De Roeck, A. and H{\"a}chler, H. P. and Hirt, A. M. and Joergensen, M. -D. and Katre, A. and Mermod, P. and Milstead, D. and Sloan, T.},
    title = "{Development of a magnetometer-based search strategy for stopped monopoles at the Large Hadron Collider}",
    doi = "10.1140/epjc/s10052-012-2212-x",
    journal = "Eur. Phys. J. C",
    volume = "72",
    pages = "2212",
    year = "2012"
}

@article{Deppisch:2019kvs,
    author = "Deppisch, Frank and Kulkarni, Suchita and Liu, Wei",
    title = "{Heavy neutrino production via $Z'$ at the lifetime frontier}",
    eprint = "1905.11889",
    archivePrefix = "arXiv",
    primaryClass = "hep-ph",
    doi = "10.1103/PhysRevD.100.035005",
    journal = "Phys. Rev. D",
    volume = "100",
    number = "3",
    pages = "035005",
    year = "2019"
}

@article{ATLAS:2023esy,
    author = "Aad, Georges and others",
    collaboration = "ATLAS",
    title = "{Search for magnetic monopoles and stable particles with high electric charges in $ \sqrt{s} $ = 13 TeV $pp$ collisions with the ATLAS detector}",
    eprint = "2308.04835",
    archivePrefix = "arXiv",
    primaryClass = "hep-ex",
    reportNumber = "CERN-EP-2023-150",
    doi = "10.1007/JHEP11(2023)112",
    journal = "JHEP",
    volume = "11",
    pages = "112",
    year = "2023"
}

\end{document}